\documentclass[aps,prd,twocolumn,a4paper,showpacs,nofootinbib,floatfix]{revtex4}
\usepackage{graphicx,amsmath,amsfonts,amssymb,array,calc,tabularx,rotating}
\usepackage{dcolumn}
\usepackage{bm}
\newlength\mylen

\def\lsim{\raise 0.4ex\hbox{$<$}\kern -0.8em\lower 0.62
ex\hbox{$\sim$}}
\def\gsim{\raise 0.4ex\hbox{$>$}\kern -0.7em\lower 0.62
ex\hbox{$\sim$}}

\begin{document}

\title{ Constraining Gravitino Dark Matter with the Cosmic Microwave
Background}
\author{Raphael Lamon}
\affiliation{Institut f\"ur Theoretische Physik, ETH Z\"urich,
  H\"onggerberg, 8093 Z\"urich, Switzerland.}
\author{Ruth Durrer}
\affiliation{D\'epartement de Physique Th\'eorique, Universit\'e de
Gen\`eve, 24 quai Ernest Ansermet, 1211 Gen\`eve 4, Switzerland.}

\date{\today}

\begin{abstract}
We consider super-gravity models in which the lightest
supersymmetric particle (LSP) is a stable gravitino. The
next-to-lightest supersymmetric particle (NLSP) freezes out with its
thermal relic density and then decays after $(10^5-10^{10})$~sec,
injecting high-energy photons into the cosmic plasma. These photons
heat up the electron plasma which then thermalizes with the cosmic
microwave background (CMB) via Compton scattering, bremsstrahlung
and double-Compton scattering. Contrary to previous studies which
assume instantaneous energy injection, we solve the full kinetic
equation for the photon number density with a source term describing
the decay of the NLSP. This source term is based on the requirement
that the injected energy be almost instantaneously redistributed by
Compton scattering, hence leading to a time-dependent chemical
potential. We investigate the case of a stau NLSP and determine the
constraints on the gravitino and stau masses from observations of
the CMB spectrum by assuming that all gravitino LSPs come from stau
NLSP decays. Unlike the analytical approximations, we find that
there may be a stau mass below which the constraint from the CMB
spectrum vanishes.
\end{abstract}

\pacs{95.35.+d, 98.80.Es, 98.80.Cq}

\maketitle

\section{\label{sec:Einleitung}Introduction}
Supersymmetry provides mainly two compelling candidates for cold
dark matter: either the gravitino or the neutralino, depending on
which one is the LSP. If we require that R-parity be conserved,
the NLSP decays into the stable LSP and releases energy in standard model
particles. At leading order, these late decays are two-body and the
accompanying energy is mainly electromagnetic.

Part of the electromagnetic release is transferred to the cosmic
microwave background radiation. The CMB then re-thermalizes through
three relevant processes:
Compton scattering $(\gamma+e\rightarrow\gamma+e)$, double-Compton
scattering $(\gamma+e\rightarrow\gamma+\gamma+e)$ and bremsstrahlung
$(e+p\rightarrow e+p+\gamma)$. The energy injection may distort the
CMB, depending on the redshift at which it occurs and on the various
time scales of the processes. Early NLSP decays can be
fully thermalized, whereas distortions caused by injection from late
decays cannot. Varying the NLSP and LSP masses, it is possible to
control both the NLSP lifetime and the energy injected in the
CMB.

The CMB is not only very isotropic, but it also has a very precise
Planck spectrum. The FIRAS instrument aboard the COBE satellite
constrains the deviation from a perfect blackbody spectrum in terms of a
few numbers. Important for this work is the limit for the chemical
potential~\cite{fixsen:96}

$$|\mu|\leq 9\times 10^{-5} ~. $$

This bound has been used to derive limits for the energy released by
NLSP decays as well as the NLSP lifetime. Observational constraints
on the CMB spectrum can be translated into bounds for the stau NLSP
and gravitino LSP masses. It should be mentioned that we explicitly
assume that all gravitinos present in the universe are produced 
by stau decays. Some models suggest that gravitino LSPs may
also be produced by scattering interactions after reheating, leading
to less stringent constraints. Recently, several papers have
employed an analytic approximation to determine these
limits~\cite{feng:0302,feng:0306,roszkowski:04}. The approximation
used in~\cite{feng:0302,feng:0306,roszkowski:04} is derived in
Ref.~\cite{hu:93:2}. This analytical result turns out to provide the
most stringent limit on the gravitino dark matter model in some
range for the NLSP and LSP masses. This prompted us to repeat the
calculation numerically.

We find that the bounds for a chemical potential of
$\mu<9\times10^{-5}$ given by \cite{hu:93:2} is a good approximation
only for stau masses above 500 GeV. Below this mass, our bounds are
less stringent and even disappear for staus lighter than 100 GeV. We
also consider the limit $\mu<10^{-5}$ and find that our results
suggest lighter gravitinos or equivalently shorter stau lifetimes.
Finally, we consider an upper bound on the chemical potential of
$2\times10^{-6}$ as planned to be achieved in the DIMES
experiment~\cite{DIMES}.  We find that, if DIMES does not see a
chemical potential, $\mu < 2\times 10^{-6}$, gravitinos cannot
significantly contribute to the dark matter if supersymmetry breaking
is gravity mediated.

After recalling some properties of the NLSP in the next section, we
write down the full kinetic equation for the evolution of the photon
one-particle distribution function in
Section~\ref{sec:evolutionequation}. We express it as an evolution
equation for a frequency-dependent chemical potential. This allows
us to determine the chemical potential for a given point in the
$(m_{\mathrm{NLSP}},m_{\mathrm{LSP}})$ and
$(m_{\mathrm{NLSP}},\tau_{\mathrm{NLSP}})$ plane of the NLSP and to
draw exclusion plots in Section~\ref{sec:CMBconstraint}. We
summarize our conclusions in Section~\ref{sec:summary}.

\section{\label{sec:NLSPproperties}NLSP properties}

 We investigate a super-gravity model with a gravitino LSP and a
stau NLSP. We assume that the NLSP freezes out with thermal
relic density and decays after a time determined by both its mass
and the LSP mass. Due to the suppression of non-photonic decay
channels, the branching ratio for decays to photons is set to be
equal to one.

Gravitino LSPs are produced through NLSP decays
$\mathrm{NLSP}\rightarrow \mathrm{LSP}+\mathrm{SMP}$, where SMP are
standard model particles. Using the standard $N=1$ super-gravity
Lagrangian, the rates for the various decay channels of the NLSP can
be calculated.

\subsection{\label{sec:sleptonNLSP}Slepton NLSP}

We assume a gravity-mediated supersymmetry breaking model, where the
gravitino LSP mass is of the order $10^2-10^4$ GeV and the stau NLSP
lifetime of the order $10^4-10^{10}$~sec.  In
gauge-mediated supersymmetry breaking models the gravitino is also
the LSP but is much lighter $(m_{\tilde{G}}\lesssim$ keV), resulting
in a much shorter NLSP lifetime. In such models, the present CMB
constraints do therefore not apply.

The width for the decay of any sfermion $\tilde{f}$ to a gra\-vi\-tino
$\tilde{G}$ for a negligible fermion mass is given by
\begin{equation}\label{sleptondecaywidth}
    \Gamma(\tilde{f}\rightarrow f\tilde{G})=\frac{1}{48\pi
    M_*^2}\frac{m_{\tilde{f}}^5}{m^2_{\tilde{G}}}\left[1-\frac{m^2_{\tilde{G}}}{m^2_{\tilde{f}}}\right]^4,
\end{equation}
where $M_*=(8\pi G_N)^{-1/2}$ is the reduced Planck mass.

Stau NLSPs decay to taus and gravitinos, which then decay to $e$,
$\mu$, $\pi^0$, 
$\pi^\pm$ and $\nu$. As mentioned in Ref.~\cite{feng:0306}, the
electromagnetic energy produced in $\tau$ decays varies between
$\epsilon^{\mathrm{min}}_{\mathrm{EM}}\approx \frac{1}{3}E_\tau$ and
$\epsilon^{\mathrm{max}}_{\mathrm{EM}}= E_\tau$. If not specified, a
value of $\epsilon_{\mathrm{EM}}=0.8 E_\tau$ will be assumed
throughout of this paper.

\subsection{\label{sec:thermalrelicdensity}Thermal relic density}
Using the thermal relic density of the right-handed slepton NLSPs
determined in Ref.~\cite{scherrer:86} and the thermally-averaged
cross section from Ref.~\cite{asaka:00}, the stau relic abundance is
given by
\begin{equation}\label{Omegastaufinal}
    \Omega_{\tilde{\tau}}^{\mathrm{th}}h^2\approx 0.2
    \left[\frac{m_{\tilde{\tau}}}{\mathrm{TeV}}\right]^2.
\end{equation}
As long as the staus do not decay, their time-dependent number density
can be expressed as
\begin{equation}\label{noft}
    n_{\tilde{\tau}}(t)\approx0.26\;
    \mathrm{m}^{-3} \left[ \frac{\mathrm{GeV}}{m_{\tilde{\tau}}}
    \right] \left[\frac{T}{\mathrm{Kelvin}}\right]^3\Omega_{\tilde{\tau}}.
\end{equation}
The fact that the final gravitino density, $\Omega_{\tilde{G}}h^2 =
(m_{\tilde G} /m_{\tilde{\tau}})\Omega_{\tilde{\tau}}h^2$, is
bounded by observations, $\Omega_{\tilde{G}}h^2\leq 0.14$ \cite{spergel:03}
together with Eq.~\eqref{Omegastaufinal} implies an upper bound for
the gravitino mass as a function of the stau mass:
\begin{equation}\label{upperboundgravmass}
    \frac{m_{\tilde{G}}}{\mathrm{GeV}} < 0.7 \times 10^6
    \frac{\mathrm{GeV}}{m_{\tilde{\tau}}}.
\end{equation}

\section{\label{sec:evolutionequation}Evolution equation for the
photon number density}
The decay of unstable particles into
photons during the early stages of the universe can lead to
distortions in the CMB. Depending on the redshift at which energy is injected,
this may leave a measurable imprint of the early
decays. This is the process which we now analyze in detail.

\subsection{\label{sec:energyinjection}Energy injection by particle
decays}

Energy injection resulting from NLSP decays heats the electrons,
leading to a ratio $T_e/T_{\mathrm{CMB}}$ larger than one, where
$T_e$ is the electron temperature and $T_{\mathrm{CMB}}$ is the
temperature of the CMB. Due to the tight coupling between electrons
and photons during the early stages of the universe, the energy
surplus of the electrons is redistributed among the photons,
distorting the CMB photon distribution from a blackbody spectrum.
Assuming that the energy transfer between electrons and
photons results in a Bose-Einstein spectrum with a
frequency-independent chemical potential, it is possible to relate
this resulting chemical potential to the number and energy density
of the injected photons and electrons.

Following the analysis done in \cite{hu:93:2}, we can write the
energy in a Bose-Einstein distribution as
\begin{equation}\label{rhoBE}
    \rho_{\mathrm{BE}}=4\sigma_{SB}T_e^4\left(1-\frac{90\zeta(3)}{\pi^4}\mu_{\mathrm{inj}}\right),
\end{equation}
where we assume a small chemical potential. The number density is
given by
\begin{equation}\label{nBE}
    n_{\mathrm{BE}}=\frac{2\zeta(3)}{\pi^2}T_e^3\left(1-\frac{\pi^2}{6\zeta(3)}\mu_{\mathrm{inj}}\right).
\end{equation}
Here $\zeta$ denotes the Riemann $\zeta$-function (see
Ref.~\cite{abramowitz:72}). 
Furthermore, due to energy conservation, we know that the energy
density may also be written as
\begin{equation}\label{rhoBE2}
    \rho_{\mathrm{BE}}=\rho_{\mathrm{P}}+\rho_{\mathrm{decay}},
\end{equation}
where $\rho_{\mathrm{decay}}$ is the energy density injected by the
NLSP decays and $\rho_P$ is the  density of the CMB photons. This
equation is only valid if the injected photons are 
redistributed in a negligible amount of time compared to the time
scales of double-Compton and bremsstrahlung. However, this
does not hold for the low-frequency spectrum, where the
photon-creating processes dominate.

More precisely, the injected
energy density is given by the following differential equation
\begin{equation}\label{ddtrhodecay}
    \frac{d\rho_{\mathrm{decay}}}{dt}=\epsilon
    \frac{m^2_{\mathrm{NLSP}}-m^2_{\mathrm{LSP}}}{2m_{\mathrm{NLSP}}}n_{\mathrm{NLSP}}(t)\frac{e^{-t/\tau}}{\tau}-4\frac{\dot{a}}{a}\rho_{\mathrm{decay}},
\end{equation}
where $\tau$ is the lifetime of the NLSP. Due to the fact that tau
decays also produce several neutrinos, the right hand side of
Eq.~\eqref{ddtrhodecay} has been multiplied by a factor $\epsilon$
describing the ratio of the injected energy to the total energy. As
pointed out in Sec.~\ref{sec:sleptonNLSP}, $\epsilon$ may have a
value between 0.3 and 1; we set $\epsilon=0.8$. In order to solve
Eq.~\eqref{ddtrhodecay}, we integrate both sides from
$t_{\mathrm{in}}$ to $t$ and obtain
\begin{eqnarray}\label{rhodecay}
    \rho_{\mathrm{decay}}&=&\epsilon
    \frac{m^2_{\mathrm{NLSP}}-m^2_{\mathrm{LSP}}}{2m_{\mathrm{NLSP}}}n_{\mathrm{NLSP}}(t_{\mathrm{in}})\left(\frac{a_{\mathrm{in}}}{a}\right)^4 \nonumber\\
    &&\times \left\{\frac{1}{2}\sqrt{\pi}\left[\mathrm{Erf}\left(\sqrt{t/\tau}\right)-\mathrm{Erf}\left(\sqrt{t_{\mathrm{in}}/\tau}\right)\right]\right. \nonumber \\
    &&\qquad-\left.\left[\sqrt{\frac{t}{t_{\mathrm{in}}}}e^{-t/\tau}-e^{-t_{\mathrm{in}}/\tau}\right]\right\},
\end{eqnarray}
where $\mathrm{Erf}$ is the error function as defined in
\cite{abramowitz:72}. Similarly, the photon number density is given
by
\begin{equation}\label{nBE2}
    n_{\mathrm{BE}}=n_{\mathrm{P}}+n_{\mathrm{decay}},
\end{equation}
where $n_{\mathrm{decay}}$ is the injected photon number density
given by the following differential equation
\begin{equation}\label{ddtndecay}
    \frac{dn_{\mathrm{decay}}}{dt}=N_\gamma n_{\mathrm{NLSP}}(t)
\frac{e^{-t/\tau}}{\tau}-3\frac{\dot{a}}{a}n_{\mathrm{decay}}.
\end{equation}
Here $N_\gamma$ is the number of photons per stau decay injected
into the spectrum. The solution of Eq.~\eqref{ddtndecay} is given by
\begin{equation}\label{ndecay}
    n_{\mathrm{decay}}=N_\gamma
    n_{\mathrm{NLSP}}(t_{\mathrm{in}})
    \left(\frac{a_{\mathrm{in}}}{a}\right)^3\left(1-e^{-t/\tau}\right).
\end{equation}

\begin{figure}
\begin{center}
  \includegraphics[width=7cm]{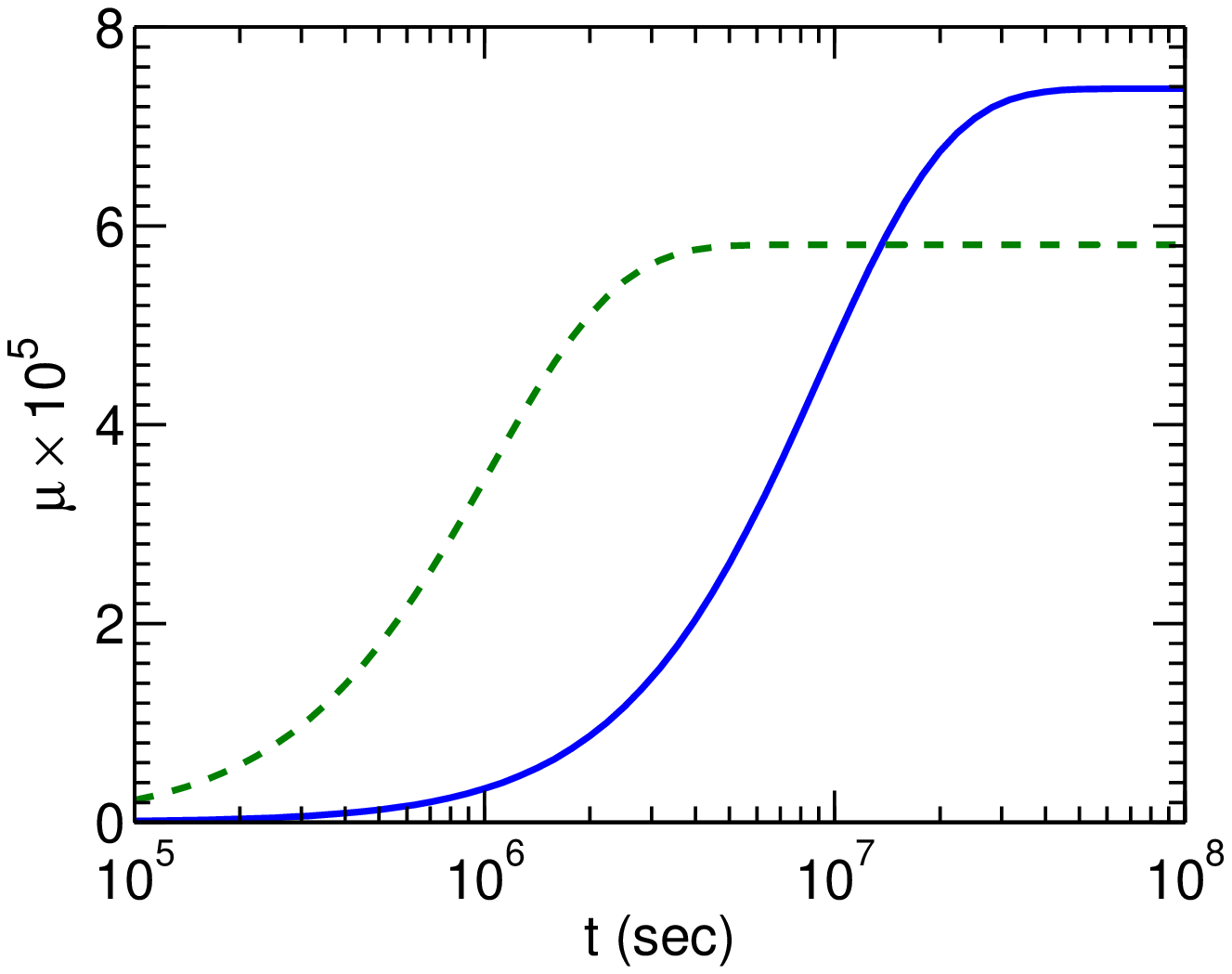}\quad
  \includegraphics[width=7cm]{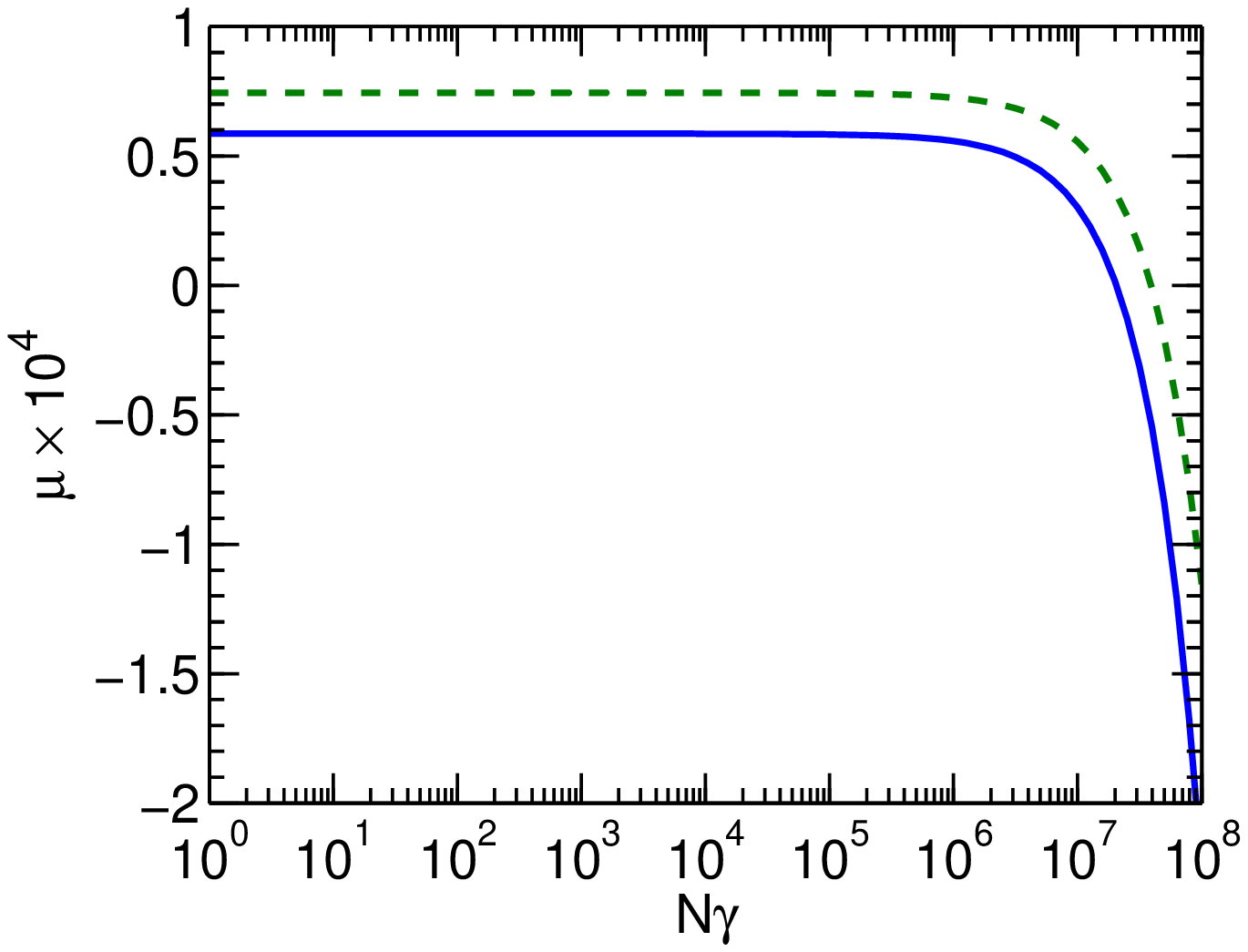}\\
  \caption{Solution of the system of equations \eqref{systemnrho}.
Top: Time evolution of the chemical potential for
    fixed $N_\gamma$ at
$m_{\tilde{\tau}}=200$ GeV and $m_{\tilde{G}}=50$ GeV, corresponding
to $\tau_{\mathrm{NLSP}}\approx
  6\times 10^6$~sec (blue solid line) and $m_{\tilde{\tau}}=300$
GeV and $m_{\tilde{G}}=50$ GeV, corresponding to
$\tau_{\mathrm{NLSP}}\approx
  7\times 10^5$~sec (green dashed line). A value of $N_\gamma=10^4$
  has been assumed.
  Bottom: The chemical potential at a time $t=10^8$~sec is shown as a function of $N_\gamma$.
  Both lines are given by the same masses as above.}\label{fig:muC}
\end{center}
\end{figure}

Inserting Eq.~\eqref{rhodecay} and Eq.~\eqref{rhoBE2} into
Eq.~\eqref{rhoBE}, as well inserting ~\eqref{ndecay} and
Eq.~\eqref{nBE2} into Eq.~\eqref{nBE} we find the relations:
\begin{eqnarray}\label{systemnrho}
  \nonumber 1+\frac{n_{\mathrm{decay}}}{n_{\mathrm{P}}} &=& \left(\frac{T_e}{T}\right)^3\left(1-\frac{\pi^2}{6\zeta(3)}\mu_{\mathrm{inj}}\right) \\
  1+\frac{\rho_{\mathrm{decay}}}{\rho_{\mathrm{P}}} &=& \left(\frac{T_e}{T}\right)^4\left(1-\frac{90\zeta(3)}{\pi^4}\mu_{\mathrm{inj}}\right).
\end{eqnarray}
These equations cannot be solved simultaneously since there are
three unknowns, $\mu_{\mathrm{inj}}$, $T_e$ and $N_\gamma$. However,
it turns out that the chemical potential is independent of $N_\gamma$ up to an
unreasonable photon number injection of $\approx10^7$ photons per
NLSP decay, as shown in the bottom panel of Fig.~\ref{fig:muC}. The
top panel shows the time evolution of the chemical potential for two
different stau lifetimes.

Up to now, we have assumed that the energy injection caused by stau
decays was instantaneously converted in a chemical potential through
Compton scattering. However, there are also two other processes that
do not conserve the photon numbers: bremsstrahlung and
double-Compton scattering. The influence of both processes is
discussed in the next section.

We shall use the value $\mu_\mathrm{inj}$ obtained in this section
as initial condition for the  numerical solution of the Boltzmann
equation discussed below.

\subsection{\label{sec:photon-matter interaction}Photon-matter interaction}
When the universe is more than a few minutes old, the coupling of
CMB photons and matter is basically due to three processes: Compton
scattering, double Compton scattering and bremsstrahlung. If these
processes are no longer very efficient, the spectrum can be
distorted. Especially, if double Compton scattering and
bremsstrahlung become weak, the photon number can no longer be
changed and energy injection into the CMB leads to a chemical
potential. With this in mind, we parameterize a general distorted
spectrum by a Bose-Einstein distribution with a frequency-dependent
dimensionless chemical potential $\mu(x,t)$,
\begin{equation}
 n(x,t)=\frac{1}{\exp(x+\mu(x,t))-1},
\end{equation}
where $x=h\nu/T_e$ is the dimensionless photon frequency. The
collision terms in the Boltzmann equation for the three relevant
processes (Compton
scattering, double-Compton scattering and bremsstrahlung) are
studied in \cite{lightman:81}. The Boltzmann equation is given by

\begin{eqnarray}
   \frac{\partial n}{\partial t}=&&\frac{1}{t_{\gamma
   e}}\frac{T_e}{m_e}\frac{1}{x^2}\frac{\partial}{\partial x}\left(\frac{\partial n}{\partial
   x}+n+n^2\right) \nonumber \\
   &&+\frac{Q g(x)}{t_{\gamma e}}\frac{1}{e^x
   x^3}\left[1-n\left(e^x-1\right)\right]\nonumber \\
   &&+\left[1-\theta(x-1)\right]\frac{1}{t_{\gamma
   e}}\frac{4\alpha}{3\pi}\left(\frac{T_e}{m_e}\right)^2\frac{1}{x^3}\nonumber\\
   &&\qquad\times\left[1-n\left(e^x-1\right)\right] \int dx x^4(1+n)n \nonumber\\
   &&-\frac{e^{x+\mu_{\mathrm{inj}}}}{\left(e^{x+\mu_{\mathrm{inj}}}-1\right)^2}\frac{d\mu_{\mathrm{inj}}}{dt}.
\label{fullmu}
\end{eqnarray}
 The first term describes Compton scattering, the second
bremsstrahlung, the third term double Compton scattering and the
fourth is the injection term given by the solution of
Eq.~\eqref{systemnrho}. We have introduced the Heaviside function
$\theta$ in the double-Compton scattering term, to take into account that
it is active only for $x<1$. The
constant $t_{\gamma e} = (n_e\sigma_T)^{-1} $ is the Thomson
scattering time, $Q = 2\sqrt{2\pi}(m_e/T_e)^{1/2}\alpha n_B T_e^{-3}
\simeq 1.7\times
10^{-10}(\mathrm{MeV}/T)^{1/2}(T/T_e)^{7/2}\Omega_Bh^2$ and $g(x)$
is the Gaunt factor. More details on the collision terms can be
found in Refs.~\cite{lightman:81,hu:93:2}
or~\cite{padmanabhan:02:vol3}.

Since we expect a small value of the chemical potential, we can expand
this equation to first order in $\mu$.
\begin{eqnarray}
    n(x,t) &\approx&
    n_0(x,t)+\mu(x,t)\frac{\partial n_0}{\partial\mu}(x,t) \nonumber \\
 &=&\frac{1}{e^x-1}-\mu(x,t)\frac{e^x}{(e^x-1)^2}.   \label{nTaylor}
\end{eqnarray}
The zeroth order is the equilibrium distribution, and the first
order in $\mu$ describes the spectral distortion. The kinetic
equations for the three relevant processes (Compton scattering,
double Compton scattering and bremsstrahlung) \cite{lightman:81},
then becomes a linear equation for the evolution of the chemical
potential $(\mu'=\partial \mu/\partial x)$,
\begin{eqnarray}
        -\frac{e^x}{(e^x-1)^2}\frac{\partial}{\partial t}\mu(x,t)=
 \frac{2}{t_{\gamma
        e}}\frac{T_e}{m_e}\frac{xe^{2x}}{(e^x-1)^4}\times \qquad
  && \nonumber \\
\left[(4-4\cosh x+x\sinh
        x)\mu'(x,t)   \right.  \left.
 -x(\cosh x-1)\mu''(x,t) \right]\negthickspace \negthinspace &&
  \nonumber\\
        +\frac{Qg(x)}{t_{\gamma
        e}}\frac{1}{x^3(e^x-1)}\mu(x,t) && \nonumber\\
        +\left[1-\theta(x-1)\right]\frac{1}{t_{\gamma
        e}}\frac{16\pi^3\alpha}{45} \left(\frac{T_e}{m_e}\right)^2
 \frac{e^x}{x^3(e^x-1)}\mu(x,t)  && \nonumber \\
  -\frac{e^x}{\left(e^x-1\right)^2}\frac{d\mu_{\mathrm{inj}}}{dt}.\nonumber\\
\label{linmu}
\end{eqnarray}

We have solved both systems of equations numerically and find
consistent results. When interested in values $\mu \sim 10^{-5}$, we
start at $t_\mathrm{in}=10^5$~sec, but when we want to detect
chemical potentials on the level of $\mu\sim 10^{-6}$ we have to
start at $t_\mathrm{in}=10^4$~sec.

\subsection{\label{sec:numevol}Time evolution of the frequency-dependent
  chemical potential}
\begin{figure}
\begin{center}
  \includegraphics[width=7cm]{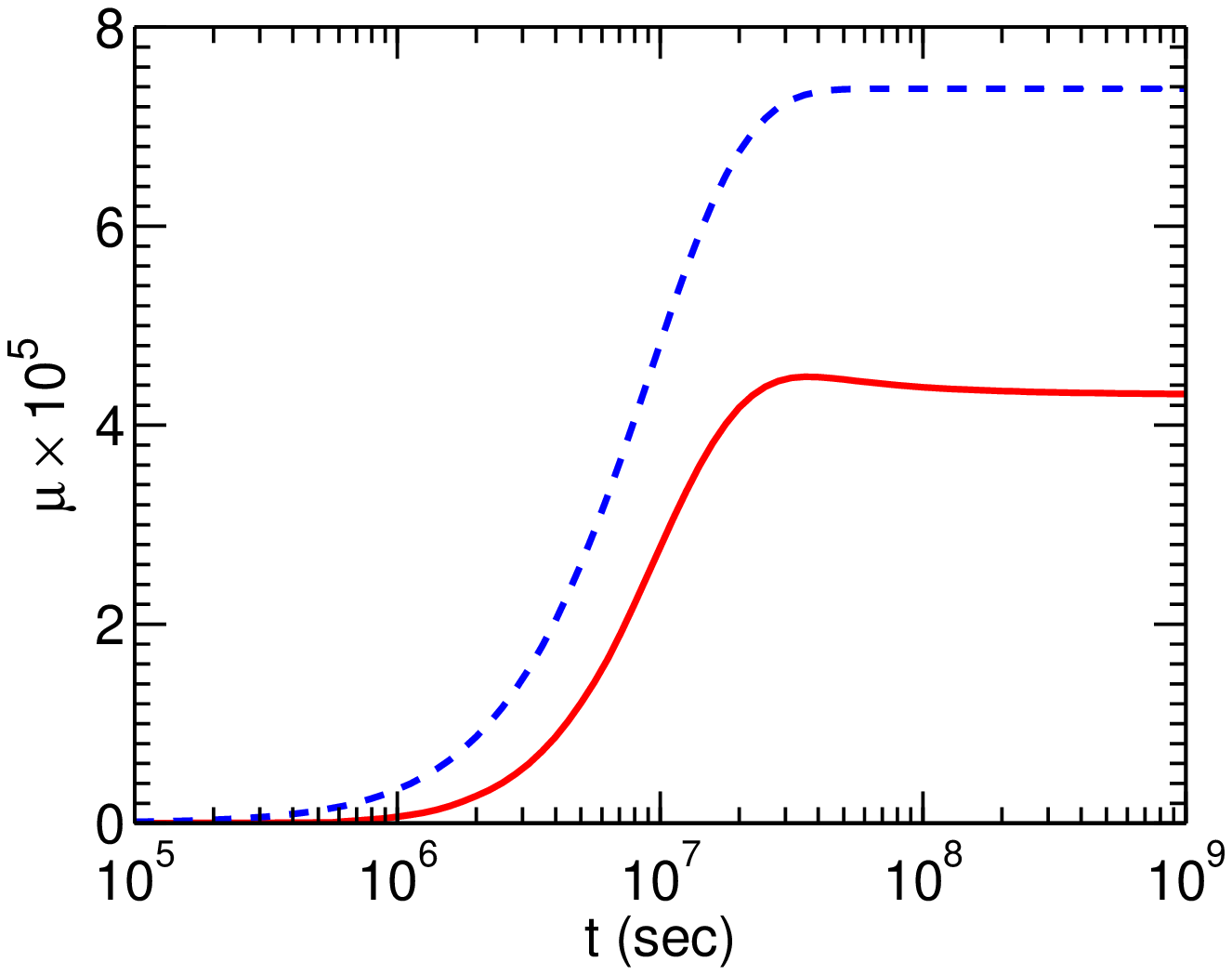}\quad
  \includegraphics[width=7cm]{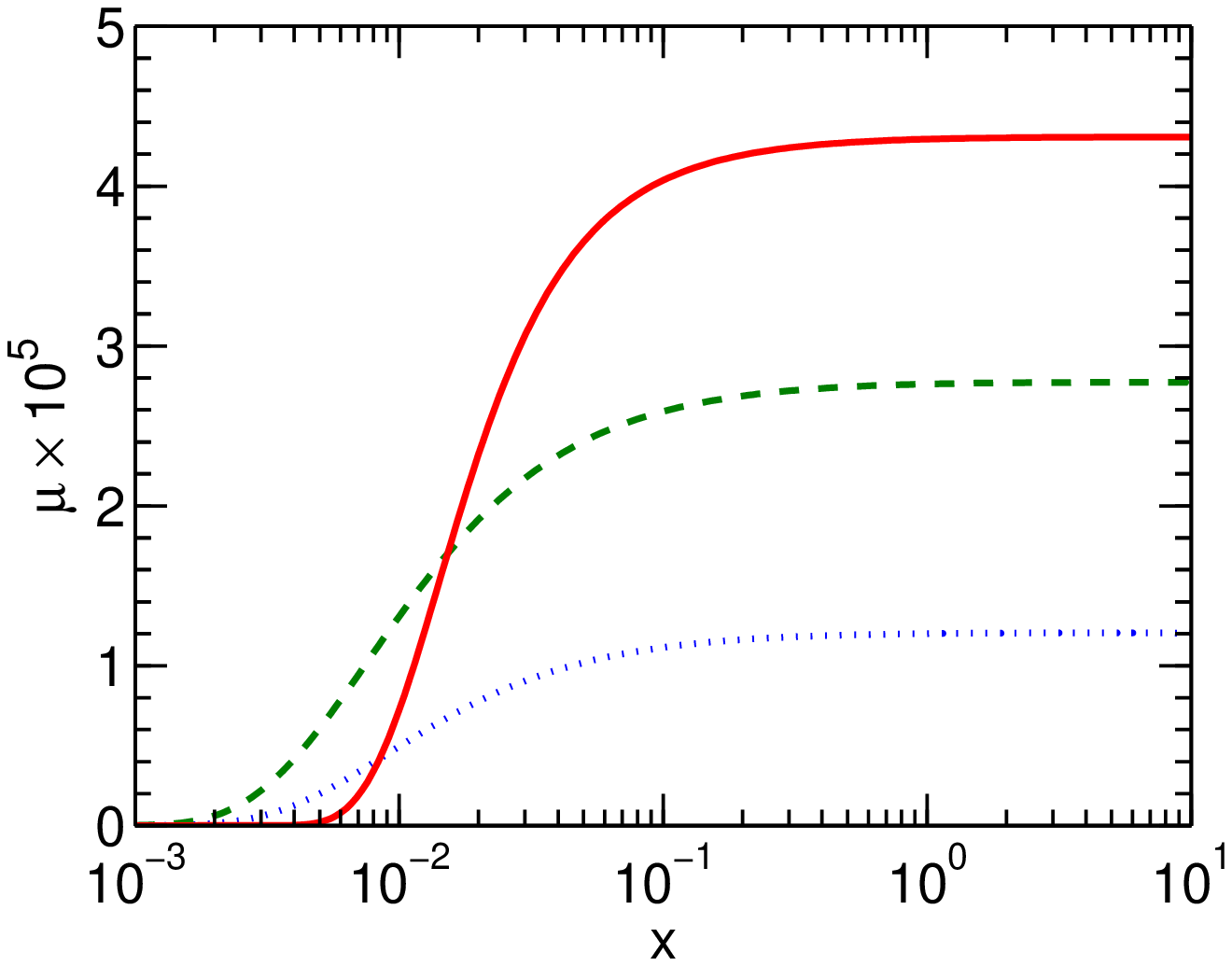}\\
  \caption{Solution of the full kinetic equation \eqref{fullmu} for $m_{\tilde{\tau}}=200$
    GeV and $m_{\tilde{G}}=50$ GeV, corresponding to
    $\tau_{\mathrm{NLSP}}\approx 6\times 10^6$~sec. Top: The dashed (blue) line
    represents the chemical potential given by
    Eq.~\eqref{systemnrho}, the solid (red) line is the solution of
    Eq.~\eqref{fullmu} evaluated at $x=3$.
  Bottom: Chemical potential as a function of $x$ given by Eq.~\eqref{fullmu} evaluated at $t=5\times10^6$~sec (dotted blue line),
  $t=10^7$~sec (dashed green line) and $t=t_{\mathrm{dec}}$ (solid red line).}\label{fig:mu}
\end{center}
\end{figure}

As shown in Ref.~\cite{padmanabhan:02:vol3}, energy injected at a
redshift higher than $z\sim 10^7$, corresponding to a time
$t\lesssim10^5$~sec is fully thermalized. Furthermore, at decoupling
time $t_{\mathrm{dec}}\approx10^{13}$~sec, the CMB spectrum is
frozen in and does not evolve anymore apart from redshifting the
photon momenta. However, as shown in Fig.~\ref{fig:mu}, the
photon-creating processes are unable to reduce a chemical potential
already as early as $t\gtrsim 10^8$~sec, much before recombination.

We see from the top panel of Fig.~\ref{fig:mu} that, compared to the
chemical potential of a Bose-Einstein spectrum (dashed blue line),
double-Compton and bremsstrahlung significantly reduce the magnitude
of the distortions from a blackbody spectrum: the chemical potential
at late times has been reduced from $7.4\times10^{-5}$ to
$4.3\times10^{-5}$. The bottom panel of Fig.~\ref{fig:mu} shows the
frequency dependence of $\mu$ evaluated at different times. The
high-frequency range is dominated by Compton scattering. The
chemical potential is constant above $x\gtrsim 1$, describing a true
Bose-Einstein spectrum. The low-energy spectrum is dominated by the
photon-creating processes which can destroy the chemical potential
and lead to a Planck spectrum below $x\lesssim4 \times10^{-3}$ at
recombination time.

It is clear that the later the energy injection, or equivalently the
later the staus decay, the weaker are the photon-creating processes
which would reduce the distortions. However, it should be
kept in mind that our equations are valid only if Compton scattering can
achieve a Bose-Einstein spectrum. By requiring that Compton
scattering be well active during stau decays, we can put an upper
limit on the stau lifetime. Following the analysis of
\cite{padmanabhan:02:vol3}, a given spectrum can only relax to a
Bose-Einstein spectrum before $t_{\mathrm{BE}}\simeq 10^9$~sec.
Therefore, the accuracy of a solution of Eq.~\eqref{fullmu} for a
stau lifetime longer than $\tau_{\mathrm{NLSP}}\sim 5\times10^8$~sec
becomes questionable and untrustworthy for
$\tau_{\mathrm{NLSP}}\gtrsim10^9$~sec.

On the other hand, after freeze-out of Compton scattering, the
injected energy resulting from stau decays cannot be scattered
downward in frequency. But there are several other processes that
could leave an imprint on the measurable CMB spectrum. For example,
photons produced during stau decays have an energy much greater than
the electron mass and can create electron-positron pairs through the
process $\gamma+\gamma\rightarrow e^++e^-$. The rate of this process
has a typical value of $\Gamma_{\mathrm{DP}}\approx10^3$
$\mathrm{sec}^{-1}$, provided $E_\gamma\gtrsim m_e^2/22T$
\cite{kawasaki:95}. Compared to the Hubble rate $H=1/2t$, we see
that this process plays an important role in heating up the
electrons. However, the raise of the electron temperature does no
longer affect the CMB spectrum when Compton scattering has already
frozen out.

We have also analyzed the fact that the true electron temperature
$T_{e,\mathrm{true}}$ is not the same as the electron temperature
$T_e$ obtained by solving Eq.~\eqref{systemnrho} due to the influence
of bremsstrahlung and double-Compton scattering which reduce the
chemical potential from the value $\mu_\mathrm{inj}$. Given the
chemical potential $\mu$ at recombination time, we can calculate
$T_e$ by inserting $\mu$ into one of the equations of the
system~\eqref{systemnrho}. We have found that $T_{e,\mathrm{true}}$
only differs from $T_e$ by $\sim10^{-4}$, and that changing the
electron temperature in Eq.~\eqref{fullmu} by such small amounts has
no effect. (This is not so surprising, as the effect is of second
order.)

\begin{figure}[h!]
\begin{center}
  \includegraphics[width=7.5cm]{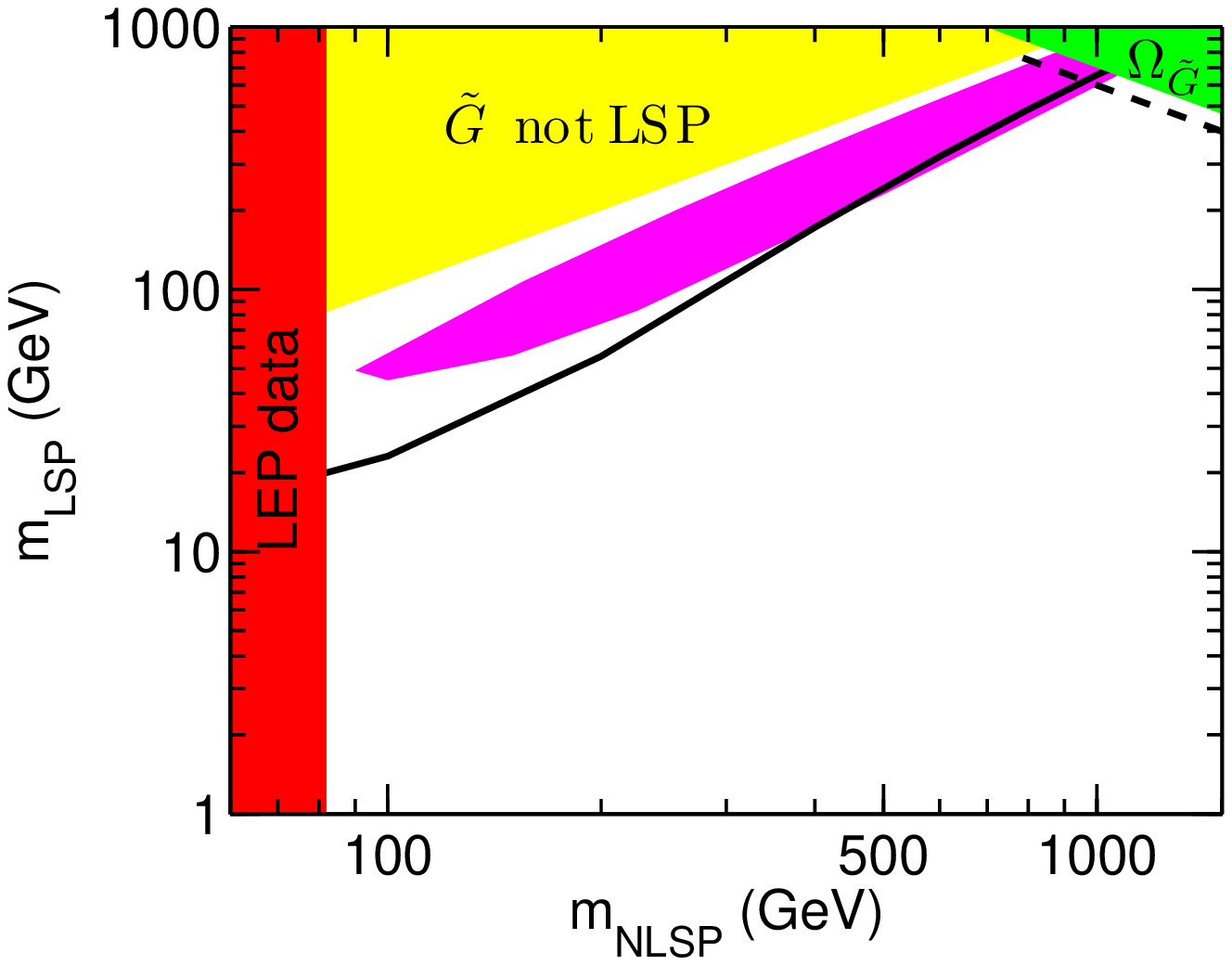}\quad
  \includegraphics[width=7.5cm]{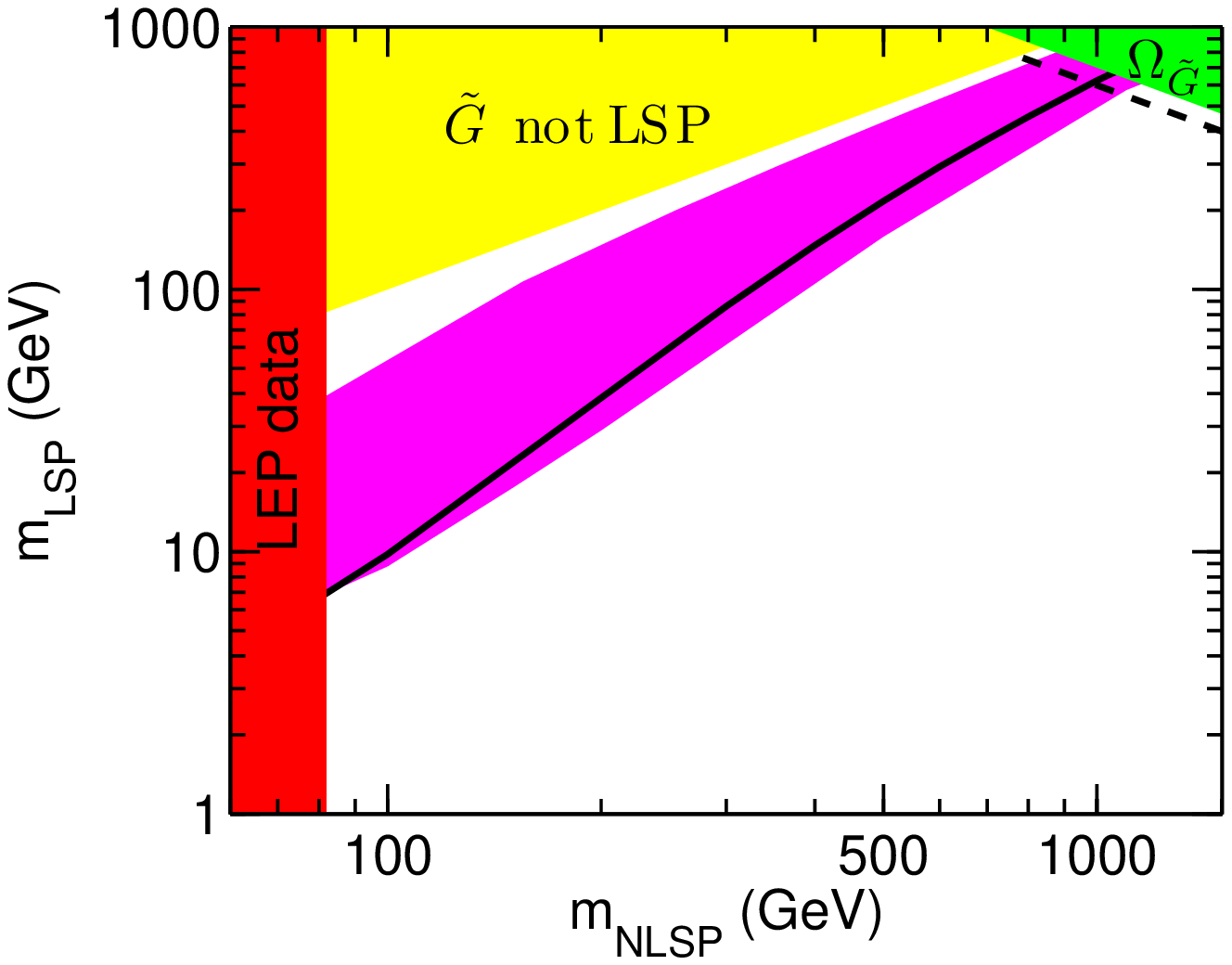}\\
  \caption{Excluded and allowed regions in the
  $(m_{\mathrm{NLSP}},m_{\mathrm{LSP}})$ parameter space. The light
  shaded region in the upper part of the graphs, labeled '$\tilde G$
  not LSP', (yellow) is not considered
  since in this part of parameter space $m_\mathrm{NLSP} <m_\mathrm{LSP}$.
  The shaded region in the right upper corner, labeled
  '$\Omega_{\tilde G}$', (green) is excluded by
  the over-closure constraint
  $\Omega_{\tilde{G}}h^2\lsim 0.14$. The dark shaded region on the
  left, labeled 'LEP data'  (red) is forbidden by LEP measurements
  \cite{pdg:04}.
   The dashed line corresponds to
  $\Omega_{\tilde{G}}h^2 = 0.12$, the best fit value for the cold dark
  matter density from the WMAP experiment~\protect\cite{spergel:03}.
  The dark shaded region in the middle (magenta) is forbidden for a
  chemical potential limit of
  $\mu <9\times10^{-5}$ (top) and $\mu<10^{-5}$ (bottom).
  The approximation of Ref.~\cite{hu:93:2}
excludes the entire
 region above the solid line.}\label{fig:CMB}
\end{center}
\end{figure}

\begin{figure}[ht]
\begin{center}
  \includegraphics[width=7.5cm]{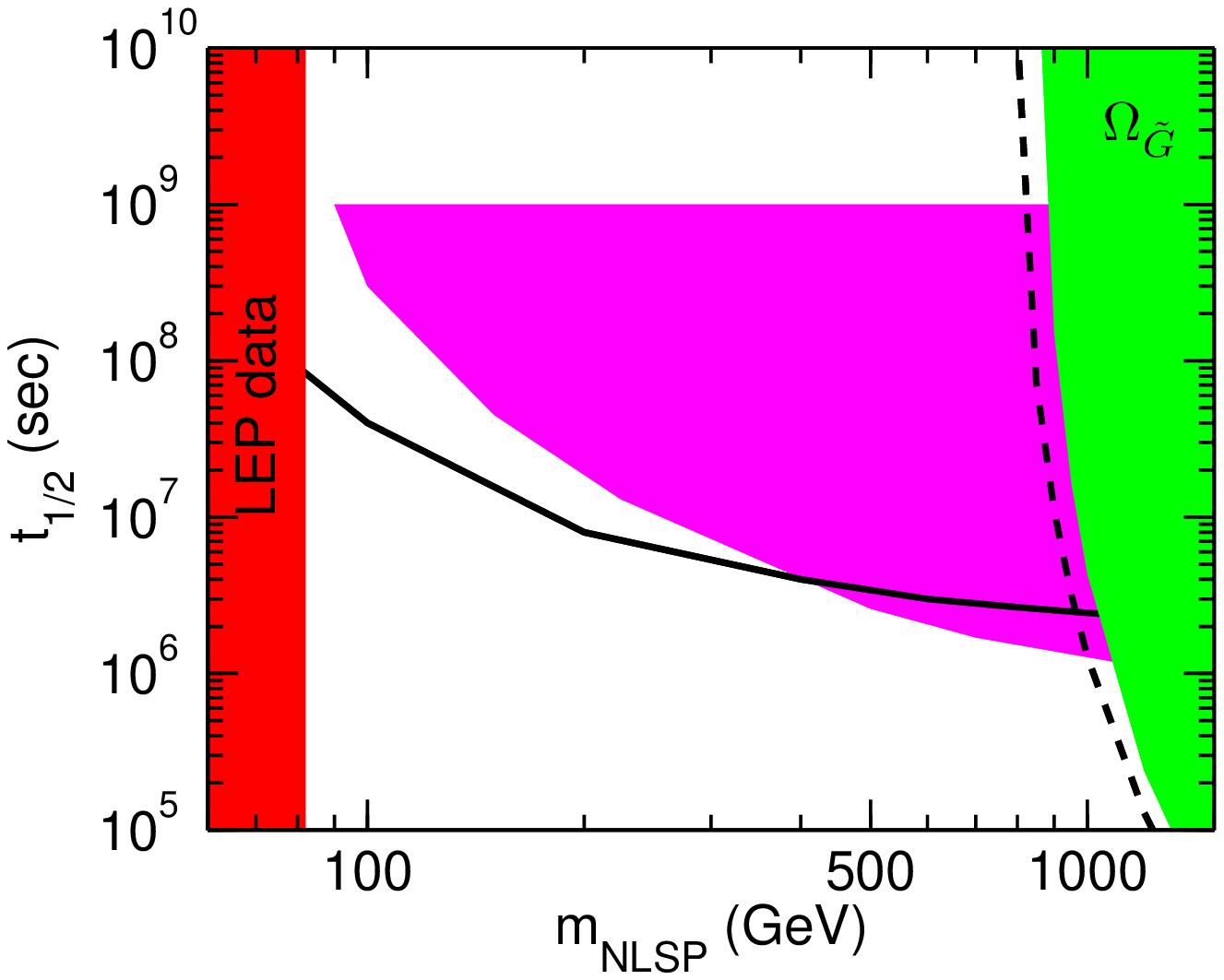}\quad
  \includegraphics[width=7.5cm]{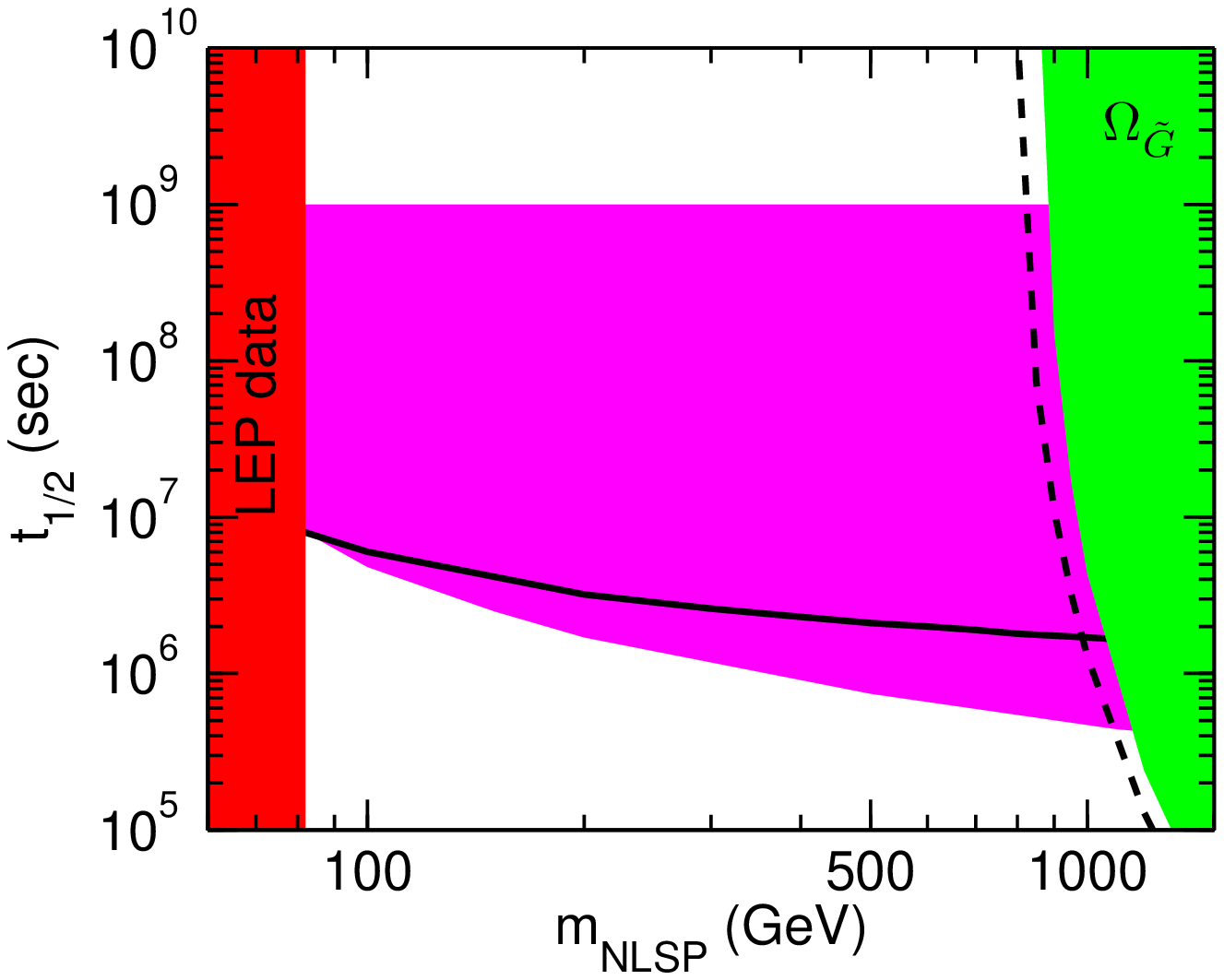}\\
  \caption{Excluded and allowed regions in the $(m_{\mathrm{NLSP}},\tau)$
  plane.
  The dashed line corresponds to
  $\Omega_{\tilde{G}}h^2 = 0.12$, the best fit value for the cold dark
  matter density from the WMAP
  experiment~\protect\cite{spergel:03}. The regions are labeled as
  in Fig.~\protect\ref{fig:CMB}. In the top panel the limit on the chemical
  potential is $\mu < 9\times10^{-5}$ while we require
  $\mu < 10^{-5}$ in the bottom panel. The approximation of Ref.~\cite{hu:93:2}
excludes the entire region above the solid line.} \label{fig:CMBlifetime}
\end{center}
\end{figure}

\begin{figure}[ht!]
\begin{center}
  \vspace{0.0cm}
  \includegraphics[width=7.5cm]{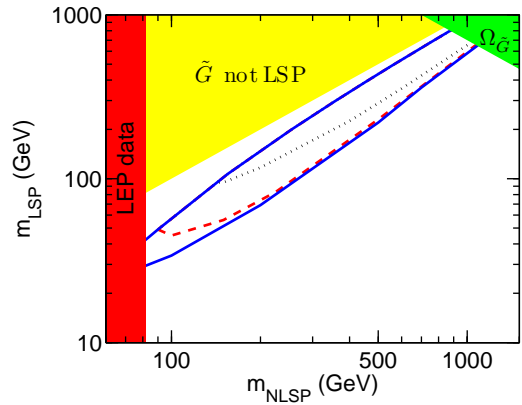}\\
  \caption{Excluded and allowed regions in the
  $(m_{\mathrm{NLSP}},m_{\mathrm{LSP}})$ plane
  for different values of the parameter $\epsilon$ and a chemical
  potential $\mu=9\times10^{-5}$. The labeled regions are as
  in Fig.~\protect\ref{fig:CMB}.
  The lower solid (blue) line represents the case $\epsilon=1$,
  the dashed (red) line $\epsilon=0.8$ and the dotted (black) line
  $\epsilon=0.3$. The upper solid line denotes the limit on the
  lifetime, $\tau>10^9$sec above which the CMB spectrum is not modified.}
 \label{fig:CMBepsilon}
\end{center}
\end{figure}

\section{\label{sec:CMBconstraint}CMB Constraints on the stau and
  gravitino masses}
The FIRAS instrument aboard the COBE satellite has measured a
 temperature $T_0 = 2.725 \pm 0.001$
Kelvin~\cite{mather:99}, and it was able to give an upper bound for
the chemical potential~\cite{fixsen:96,smoot:97},
 $|\mu|<9\times10^{-5}$.
This bound comes from measurements in the frequency range from 2 to
600 GHz, corresponding to $x=h\nu/T_0 \in [0.03 , 10]$.
 There are also some measurements at lower frequencies, but their
 accuracy is worse, leading to a lower bound on $\mu$ which is by at
 least an order of magnitude higher. To
obtain good accuracy in the measured interval, we numerically
compute the chemical potential for
 $x \in[ 10^{-4}, 15]$. We require that
 the chemical potential be never higher
than $9\times10^{-5}$ within the \emph{experimental} range $x \in
[0.03,\,10]$. Outside that range, $\mu$ may be larger (experiments do
not rule out deviations outside the frequency range
$[0.5\,\mathrm{GHz},600 \,\mathrm{GHz}]$).

A point in the $(m_{\mathrm{NLSP}},m_{\mathrm{LSP}})$-plane is
considered to satisfy the CMB observational bound if the magnitude
of the chemical potential never trespasses the limit
$9\times10^{-5}$ within and only within the frequency range
$[0.03,\,10]$. Due to the limitations explained in the previous section,
not every point in the $(m_{\mathrm{NLSP}},m_{\mathrm{LSP}})$-plane
can be calculated, but we expect the chemical potential to be much
smaller than the experimental limit for points where our calculation
cannot be trusted.

\begin{figure}[ht!]
\begin{center}
  \vspace{0.0cm}
  \includegraphics[width=7.5cm]{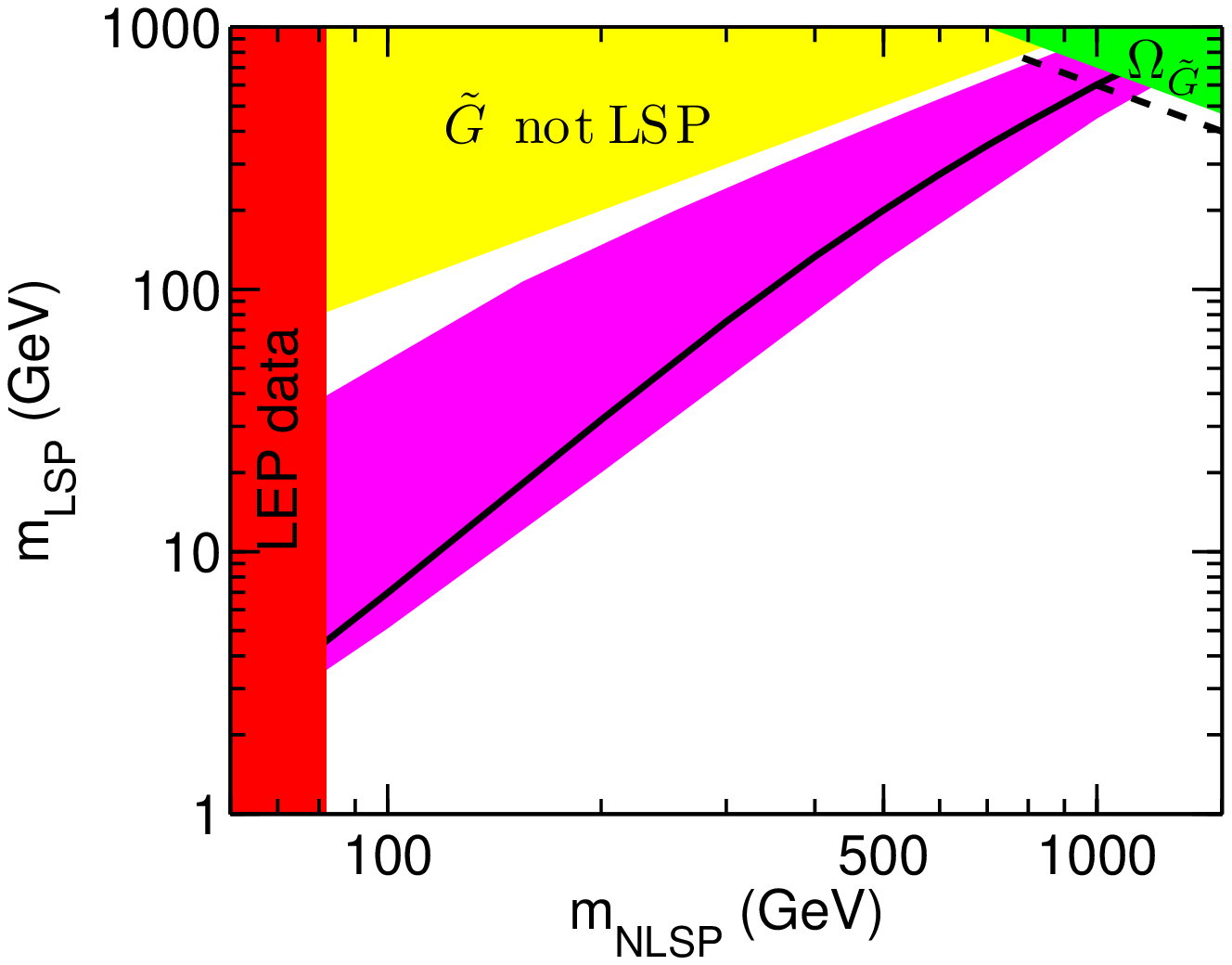}\\
  \includegraphics[width=7.5cm]{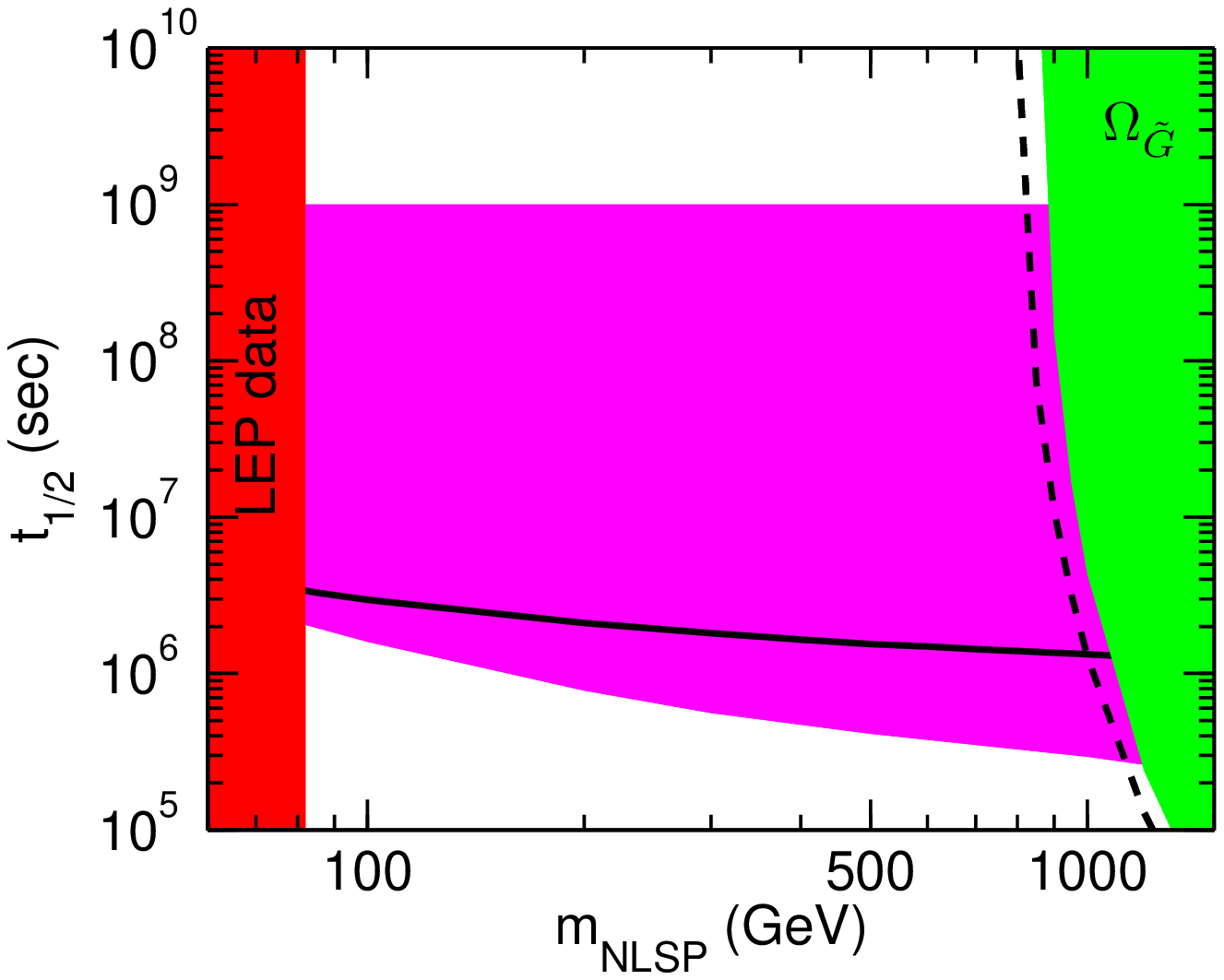}
  \caption{Excluded and allowed regions in the
  $(m_{\mathrm{NLSP}},m_{\mathrm{LSP}})$
  plane (top panel) and in the $(m_{\mathrm{NLSP}},\tau)$ plane (bottom
  panel)  for a limit on the chemical
  potential of $\mu<2\times10^{-6}$. The labeled regions are like in
  Fig.~\protect\ref{fig:CMB}.} \label{fig:DIMES}
\end{center}
\end{figure}

An estimate of the chemical potential caused by an instantaneous
energy injection is given in \cite{hu:93:2}. It is shown as a solid
line in Figs.~\ref{fig:CMB} to~\ref{fig:DIMES}. This approximation
is not very precise for small distortions. While it is in good
agreement with our results for staus heavier than 500 GeV, it does
not take account of the fact that light staus do not inject enough
energy to significantly distort the spectrum. As shown in the upper
panel of Fig.~\ref{fig:CMB}, the bound on the gravitino mass
disappears for staus lighter than $\approx100$ GeV. Moreover, due to
the freeze-out of Compton scattering, we have introduced a limit on
the gravitino masses corresponding to a stau lifetime $10^9$~sec.
All gravitino masses leading to longer stau lifetimes are allowed.

Our results match quite well with the approximation given in
\cite{hu:93:2} for $\mu<10^{-5}$, as shown in the bottom panel of
Fig.~\ref{fig:CMB}. We find the same limit for light staus and a
somewhat more stringent limit for heavy staus. However, when compared to the
stau lifetime (see the bottom panel of
Fig.\ref{fig:CMBlifetime}), our numerical results give a limit on the
lifetime that is up to five times shorter than the one obtained by
using the approximation.

As mentioned in Sec.~\ref{sec:energyinjection}, the injected energy
depends on the energy going into neutrinos. This is described by the
parameter $\epsilon$. Contrary to what is claimed in
\cite{feng:0306}, our results depend significantly on this
parameter. We considered the two extreme cases $\epsilon=0.3$ and
$\epsilon=1$ and the usual case $\epsilon=0.8$
(Fig.~\ref{fig:CMBepsilon}). The case $\epsilon=0.3$ is much less
constraining than $\epsilon=0.8$, even completely disappearing for
staus lighter than 140 GeV. On the other hand, the cases
$\epsilon=1$ and $\epsilon=0.8$ match well down to $m_{\tilde{\tau}}
\simeq 200$GeV below which the former becomes more stringent.

Future missions like the Absolute Radiometer for Cosmology,
Astrophysics Diffuse Emission (ARCADE)~\cite{arcade,kogut:04} or the
Diffuse Microwave Emission Survey (DIMES)~\cite{DIMES} experiments
may improve sensitivities in the poorly studied
centimeter-wavelength band, improving the limit on the chemical
potential to about $|\mu|<2\times10^{-6}$. In our model, if neither
DIMES nor ARCADE is able to measure distortions of the CMB,
gravitinos could only contribute to the missing dark matter if
$\tau_{\mathrm{NLSP}}\gtrsim 10^9$~sec or
$\tau_{\mathrm{NLSP}}\lesssim 2\times10^5$~sec (see
Fig.~\ref{fig:DIMES}). However, combining our results with other
constraints~\cite{feng:231}, we find that gravitinos could not
significantly contribute to the dark matter for such a bound.

\section{\label{sec:summary}Summary}
We have studied the effect on the CMB from stau NLSP decays into
gravitino LSPs, assuming that the staus freeze out with their
thermal relic density. We have numerically solved the kinetic
equation for the photon number density with non-instantaneous energy
injection. We have found that our numerical results are in good
accordance with the analytical approximation \cite{hu:93:2} for the
induced chemical potential $\mu<9\times10^{-5}$  if the stau is
heavier than 500GeV, but differs considerably for lighter stau
masses. For light staus the constraints are weaker and even
disappear for $m_{\tilde{\tau}} < 100$GeV. On the other hand, the
approximation underestimates the limits for stronger constraints
given by  $\mu<10^{-5}$ or even more for $\mu<2\times 10^{-6}$. This
limit, which could be achieved in planned
experiments~\cite{DIMES,arcade}, together with other
contraints~\cite{feng:231} would completely exclude the 
gravitino as dark matter candidate in models with gravity-mediated
supersymmetry breaking. However, allowing a gravitino production
after reheating leads to less stringent constraints than our
results. We also found that the results depend sensitively on the
energy injection parameter $\epsilon$.

\acknowledgments{We are grateful to Leszek Roszkowski for many helpful remarks
 and to J\"urg Fr\"ohlich for discussions. We also thank Choi Ki-Young
 for pointing out an important error in
 the first version of this paper. This work is supported by the Fonds National
 Suisse.

\end{document}